\documentclass[pra,twocolumn,amsmath,amssymb,aps,longbibliography,showpacs]{revtex4-1}
\usepackage{dcolumn}
\usepackage{bm}
\usepackage{CJK}
\usepackage{color}
\usepackage{makecell}
\usepackage{graphicx}

\begin{document}


\title{Domain formation and universally critical dynamics through phase separation in two-component  Bose-Einstein condensates}

\author{Yikai Ji$^{1}$, Xizhou Qin$^{1}$, Bin Liu$^{1}$, Yongyao Li$^{1}$, Bo Lu$^{2,3}$,  Xunda Jiang$^{1,2,}$}
\email{jxd194911@163.com}
\author{Chaohong Lee$^{2,3}$}
\address{$^1$School of Physics and Optoelectronic Engineering, Foshan University, Foshan 528000, China}
\affiliation{$^{2}$Institute of Quantum Precision Measurement, State Key Laboratory of Radio Frequency Heterogeneous Integration, Shenzhen University, Shenzhen 518060, China}

\affiliation{$^{3}$College of Physics and Optoelectronic Engineering, Shenzhen University, Shenzhen 518060, China}


\begin{abstract}
  We explore the defect formation and universally critical dynamics in two-dimensional (2D) two-component Bose-Einstein condensates(BECs) subjected to two types of potential traps: a homogeneous trap and a harmonic trap.
  We focus on the non-equilibrium universal dynamics of the miscible-immiscible phase transition with both linear and nonlinear quenching types.
  Although there exists spatial independence of the critical point, we find that the inhomogeneity of trap doesn't affect the phase transition of system and the critical exponents can still be explained by the homogeneous Kibble-Zurek mechanism.
  By analyzing the Bogoliubov excitations, we establish a power-law relationship between the domain correlation length, the phase transition delay, and the quench time.
  Furthermore, through real-time simulations of phase transition dynamics, the formation of domain defects and the delay of phase transition in non-equilibrium dynamics are demonstrated, along with the corresponding universal scaling of correlation length and phase transition delay for various quench time and quench coefficients, which align well with our analytical predictions.
  Our study confirms that the universality class of two-component BECs remains unaffected by dimensionality, while the larger nonlinear coefficients effectively suppress non-adiabatic excitations, offering a novel perspective for addressing adiabatic evolution.

\end{abstract}

\maketitle

\section{INTRODUCTION}
The phenomenon of spontaneous symmetry breaking (SSB) and its dynamics during phase transitions have been extensively studied in various fields~\cite{Sachdev2011,Morikawa1995,Kibble1980}, such as condensed matter physics~\cite{Ruutu1996,Bauerle1996,Monaco2009}, cosmology~\cite{Kibble1976}, and ultracold atomic gases~\cite{Anderson2008,Zurek2009,Navon2015,Lamporesi2013,Anquez2016,Clark2016,Feng2018}. One significant consequence of SSB transitions is the nucleation of topological defects, which can be effectively described by the Kibble-Zurek mechanism (KZM)~\cite{Kibble1976,Kibble1980,Zurek1985,Zurek1996,JDziarmaga2000,Polkovnikov2011,Bloch2008}. Initially proposed in the early universe, the KZM has been subsequently generalized to include condensed matter physics and other systems. The formation of defects and their universality in helium superfluids and atomic BECs have attracted considerable attention due to the analogous SSB behavior between the universe and these quantum systems.

Ultracold atomic systems, characterized by their high controllability, robust quantum coherence, and isolation from the environment, provide an excellent platform for investigating SSB phase transitions.
During a quench across a phase transition into a symmetry-broken phase, the dynamics result in the formation of defects, each independently selecting an order parameter associated with symmetry breaking.
 Experimental studies utilizing single-component atomic BECs have demonstrated the spontaneous emergence of defects, such as vortices~\cite{Anderson2008}, dark solitonic vortices~\cite{Lamporesi2013,Donadello2014}, and persistent currents~\cite{Corman2014} during the BEC phase transition, these experiments have offered quantitative validation of the KZ scaling.  Recent efforts have focused on creating multi-component BECs using different atomic species~\cite{McCarron2011,Modugno2002}, isotopes~\cite{Wieman2008}, or spin states~\cite{Lin2011,Tojo2010}. Multi-component BECs exhibit a wealth of physics that is not accessible in single-component BECs, including phase separation accompanied by symmetry breaking~\cite{Timmermans1998, Ao1998, Trippenbach2000, Esry1998, Lee2009, Alexandrov2002, Takeuchi2013}, Josephson oscillations~\cite{Williams1999}, and domain walls~\cite{Davis2011, Davis2012, Swislocki2013, Hofmann2014, Wu2017, Xu2016,Ye2018}. Notably, phase separation has been observed in several experiments involving multi-component BECs~\cite{Wieman2008, McCarron2011, Lin2011, Modugno2002, Tojo2010}. By controlling the intra- and inter-component interactions through Feshbach resonance~\cite{Courteille1998, Cornish2000, Inouye1998}, multi-component BECs provide an ideal test bed for investigating the non-equilibrium physics of phase separation.

Based on two-component BECs, Ref.~\cite{ Lee2009} had investigated the universality and symmetry-breaking transitions in coupled systems using both mean-field and full quantum frameworks. Recent studis Ref.~\cite{Davis2011, Davis2012} focus on the miscibility-immiscibility(MI) phase transition dynamics and defects formation in binary coupled BECs, taking into account the influence of homogeneous and harmonic traps on KZM scaling. It is observed that the universal scaling in a harmonic trap can be accurately described by the inhomogeneous KZM. Additionally, the paper examines the universal real-time dynamics in an array of coupled binary atomic BECs confined in optical lattices~\cite{Xu2016}. Previous studies demonstrate that the critical exponents in binary coupled BECs belong to the same universal class $(z=1, v=1/2)$. However, recent researches reveal that the MI transition induced by quenching the atom-atom interaction differs from the transition driven by quenching the Rabi coupling, belonging to distinct universal classes~\cite{Jiang2019,Jiang2023}. Specifically, two-component BECs without Rabi coupled interaction exhibit critical exponents of $(z=2, v=1/2)$. Further investigations confirm that the critical exponents are not significantly affected by nonlinear quenches during the phase transition, and a larger nonlinear coefficient can suppress systematic excitation~\cite{Jiang2023}. Despite the abundance of research on the MI phase transition in one-dimensional systems, limited attention has been given to high-dimensional systems. Recent studies explore universal coarsening dynamics and defect formation in 2D binary BECs subjected to sudden quenching of system parameters~\cite{Hofmann2014,Williamson2016,WilliamsonPRA2016}. However, the universal non-equilibrium dynamics resulting from a slow quench process remain unexplored.

In this paper, we investigate the universal non-equilibrium dynamics in 2D two-component BECs subjected to different potential traps. By quenching the interaction strength across the critical point, the formation of defect  and universal dynamics of the MI phase transition are studied.
Through theoretical analysis and numerical simulations, we establish a power-law scaling relationship between the correlation length of domains, the phase transition delay, and the quench time. Critical exponents obtained from simulations align well with theoretical predictions. Furthermore, we find that the scaling between phase transition delay and quench time remains unchanged by trap frequency, as the critical point is independent of trap inhomogeneity. This confirms that nonlinear power-law quenching and harmonic traps do not alter the critical exponent of the phase transition.

The paper is organized as follows.
In Sec.~\uppercase\expandafter{\romannumeral2}, we present the physical model and the KZM. In Sec.~\uppercase\expandafter{\romannumeral3}, we analyze the Bogoliubov excitations and analytically extract the critical exponents.
In Sec.~\uppercase\expandafter{\romannumeral4}, we present real-time nonequilibrium universal dynamics and demonstrate the formation of defects in homogeneous trap and harmonic trap. Finally, we give a brief summary and discussion in Sec.~\uppercase\expandafter{\romannumeral5}.

\section{Model}
Considering two-component BECs confined in a two-dimensional harmonic trap, which can be effectively described by the Gross-Pitaevskii equations(GPE)(we have set $\hbar=m=1$),
\begin{small}
\begin{equation*}\label{TGPE1}
i\frac{{\partial {\psi _j}}}{{\partial t}} = \left[ { - \frac{1}{2}\nabla^2+ V\left( {x,y} \right) + {g_{jj}}{{\left| {{\psi _j}} \right|}^2} + {g_{12}}{{\left| {{\psi _{3 - j}}} \right|}^2}} \right]{\psi _j},
\end{equation*}
\end{small}
where ${j = 1,2} $ denotes the distinct components of BECs, $g_{jj}$ and $g_{12}$ represent the intra-component and inter-component interactions, respectively, which can be adjusted using the Feshbach resonance. $\psi_{j}$ is the wave function for different BECs, which is normalized as follows
\begin{equation}\label{Normalization}
\int\int {{{\left| {{\psi _j}(x,y)} \right|}^2}dxdy}  = {N_j}.
\end{equation}
Generally, we have $N_{1} = N_{2} = N/2$, where $N$ is the total number of particles. The harmonic trap potential can be expressed as
 \begin{equation}\label{Harmonic}
V\left( {x,y} \right) = \frac{1}{2}{\omega ^2}\left( {{x^2} + {y^2}} \right).
\end{equation}
The competition between the intra-component interaction $g_{jj}$ and inter-component interaction $g_{12}$ gives rise to two distinct phases: the immiscible phase and the miscible phase~\cite{Timmermans1998,Ao1998,Trippenbach2000,Davis2011,Davis2012}. The phase transition between these phases is determined by the condition $g_{12}^{2}=g_{11}g_{22}$. In the case of strong intra-component interaction, where $g_{12}^{2}<g_{11}g_{22}$, the two-component BECs prefer to coexist everywhere, resulting in lower energies for the whole system. Conversely, for strong inter-component interactions, where $g_{12}^{2}>g_{11}g_{22}$, the two-component BECs prefer to be spatially separated. It is worth mentioning that the critical point remains independent of spatial variation, even in harmonic traps.

To investigate the universal dynamics of the system, we define a dimensionless distance
\begin{equation}\label{Dim_g_1}
\epsilon \left( t \right) = \left| {{g_{12}}\left( t \right) - g_{12}^c} \right|/g_{12}^c,\quad g_{12}^{c}=\sqrt{g_{11}g_{22}}.
\end{equation}
Subsequently, we perform a quench of the inter-component interaction $g_{12}$ across the critical point $g_{12}^c$, giving
\begin{equation}\label{Non_Quen_1}
 {g_{12}}\left( t \right) = g_{12}^c\left( {1 + sgn\left( t \right){{\left| {\frac{t}{{{\tau _Q}}}} \right|}^{\alpha}}} \right).
\end{equation}
Here, $\tau_Q$ represents the quench time that characterizes the speed at which the system crosses the critical point. The quench coefficient is denoted by ${\alpha}$ and $sgn$ refers to the signum function. According to KZ argument, near freeze time $\hat{t}$, the system becomes adiabatic and related $\hat{\epsilon}$ is given
\begin{equation}\label{Free_time}
 \hat{\epsilon} \sim \left({\alpha}/\tau_{Q}\right)^{\frac{{\alpha}}{1+{\alpha}vz }}, \quad \hat t = {\left( {{\alpha}{\tau _Q}^{{\alpha}vz}} \right)^{\frac{1}{{{\alpha}vz + 1}}}},
\end{equation}
the correlation length and density of excitations that undergoes freezing at $\hat{t}$ are characterized by
\begin{equation}\label{Correl_len_1}
\hat{\xi} \sim \tau_{Q}^{\frac{{\alpha}v}{1+{\alpha}vz}}, \quad  n_{ex} \simeq \hat{\xi}^{-d} \sim \tau_{Q}^{-\frac{{\alpha}vd}{1+{\alpha}vz}},
\end{equation}
where $d$ is the number of space dimensions.
\section{BOGOLIUBOV EXCITATION AND CRITICAL EXPONENTS}
For a homogeneous trap, the translation symmetry is preserved and momentum $\bf{k}$ is conserved. In the miscible phase, the nonlinear Schr$\ddot{o}$dinger equation has a clear homogeneous solution given by $\rho_{j}=|\phi_{j}|^{2}=N_{j}/L_{x}L_{y}$, where $\phi_{j}$ represents the ground state wave-function of component $j$ and $L_{x,y}$ are the lengths of the system in the $x$ or $y$ direction. The chemical potentials can be expressed as $\mu_{1}=g_{11}\rho_{1}+g_{12}\rho_{2}$ and $\mu_2=g _{22}\rho_2+g _{12}\rho_1$.

To obtain the Bogoliubov excitation spectrum, the perturbed ground state is considered as
 \begin{equation}
 \psi_{j}\left(\bf{r},t \right)=[\phi_{j}\left(\bf{r} \right)+\delta \phi_{j}\left(\bf{r},t\right)]e^{-i\mu_{j}t},   \label{Bogoliubov form1}
 \end{equation}
 where ${\bf{r}} = \left( {{x},{y}} \right)$, and the perturbations $\delta\phi_{1,2}\left(\bf{r},t\right)$ can be written as
\begin{equation}
\left( {\begin{array}{*{20}{c}}
{\delta {\phi _1}}\left(\bf{r},t\right)\\
{\delta {\phi _2}}\left(\bf{r},t\right)
\end{array}} \right) = \left( {\begin{array}{*{20}{c}}
{{u_{1,\bf{k}}}}\\
{{u_{2,\bf{k}}}}
\end{array}} \right){e^{i\bf{k}\bf{r} - i\varpi t}} + \left( {\begin{array}{*{20}{c}}
{{v_{1,\bf{k}}^{*}}}\\
{{v_{2,\bf{k}}^{*}}}
\end{array}} \right){e^{-i\bf{k}\bf{r} + i\varpi t}}.  \label{fluctuation form1}
\end{equation}
Here, ${\bf{k}} = \left( {{k_x},{k_y}} \right)$  represents the excitation quasimomentum in the $x$ or $y$ direction, and $\varpi$ is excitation frequency, respectively. $u_{j,\bf{k}}$ and $v_{j,\bf{k}}$, $(j=1,2)$ are the Bogoliubov amplitudes. Similar to the previous work~\cite{Timmermans1998,Ao1998,Trippenbach2000,Jiang2019}. One can obtain the excitation spectrum
as
\begin{equation}
\varpi_{\pm}^{2}=\bf{\epsilon_{0}}\left(\bf{\epsilon_{0}}+2\eta_{\pm} \right),    \label{excitation}
\end{equation}
where $\bf{\epsilon_{0}}=\bf{k}^2/2$, and
\begin{equation}
\eta_{\pm}=\frac{\rho}{4}\left(g_{11}+g_{22}\pm \sqrt{\left( g_{11}-g_{22}  \right)^{2}+4g_{12}^{2}}   \right).
\end{equation}
The Bogoliubov excitations in a 2D system exhibit similarities to those in a 1D system, with the only distinction being the single particle energy $\bf{\epsilon_{0}}$. By evaluating the sound velocity and the energy of the rapidly unstable mode, one can derive the critical exponents of the system as $\left(v=1/2, z=2\right)$~\cite{Jiang2019}.

\section{Defects Formation and Universality in Two Component BECs}
In this section, we investigate the non-equilibrium universal dynamics of the MI phase transition in two-component BECs under both homogeneous and harmonic potentials. By quenching the interaction across the critical point, we explore the formation of defects and delay in phase transition.
\subsection{Homogeneous Case}
In this part, we consider a scenario where the system is located in a homogeneous trap with $\omega=0$. To study its non-equilibrium dynamics, we first prepare an initial state for the system in which both components of BECs occupy all spatial space homogeneously while adding Gaussian noise to simulate quantum fluctuations. We then quench $g_{12}(t)$ across the critical point for various coefficients ${\alpha}$ and quench time $\tau_Q$ according to Eq.\ref{Non_Quen_1}. In our numerical simulation, we choose a total particle number of $N=1\times 10^{5}$, dimensionless intra-component interactions of $g_{11}=g_{22}=1$, and dimensionless length of $L_{x}=L_{y}=64$.

During this non-equilibrium process, relaxation time diverges near the critical point due to vanishing energy gap resulting in instantaneous states that cannot adiabatically follow changes in Hamiltonian. According to adiabatic-impulse-adiabatic approximation, these instantaneous states freeze or stop evolving within impulse regions separated by $\pm \left(\hat{t}\right)$. When quench parameters are outside impulse regions, states restart their evolution. However, frozen states at $-\hat{t}$ are no longer eigenstates of Hamiltonian at $+\hat{t}$. Therefore non-adiabatic defects inevitably generate defect density obeying KZ scaling for different quench times.

\begin{figure}[!htp]
   \centering\includegraphics[width=0.48\textwidth]{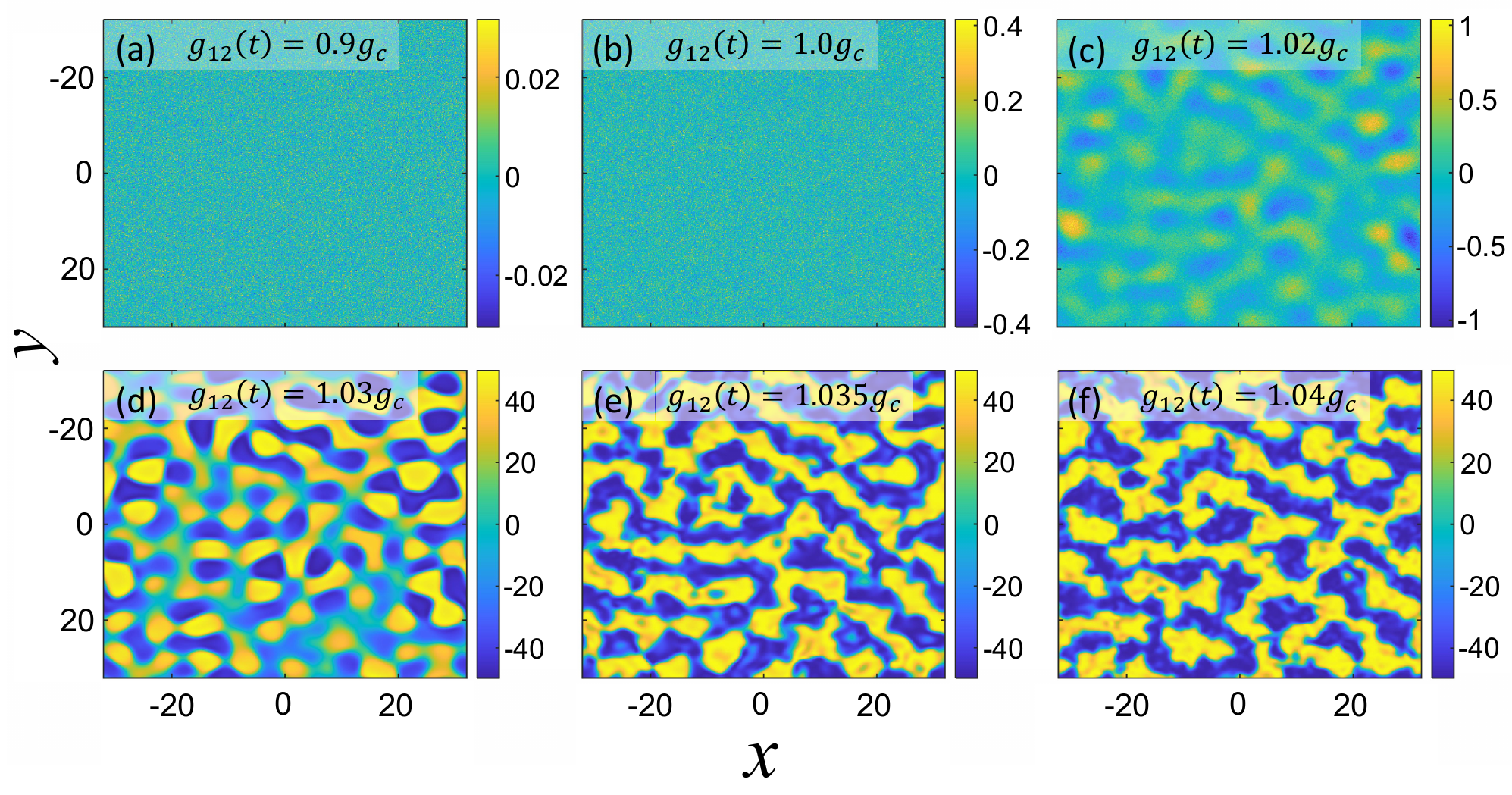}
  \caption{(Color online) Time evolution of the density difference $J(x,y)$  of two-component BECs in one of quenched dynamics. The related parameters are chosen as $\tau_Q=10^3$, $\omega=0$, and $\alpha=1$. }
\label{Fig_Dom_Form}
\end{figure}
\begin{figure}[!htp]
  \centering\includegraphics[width=0.49\textwidth]{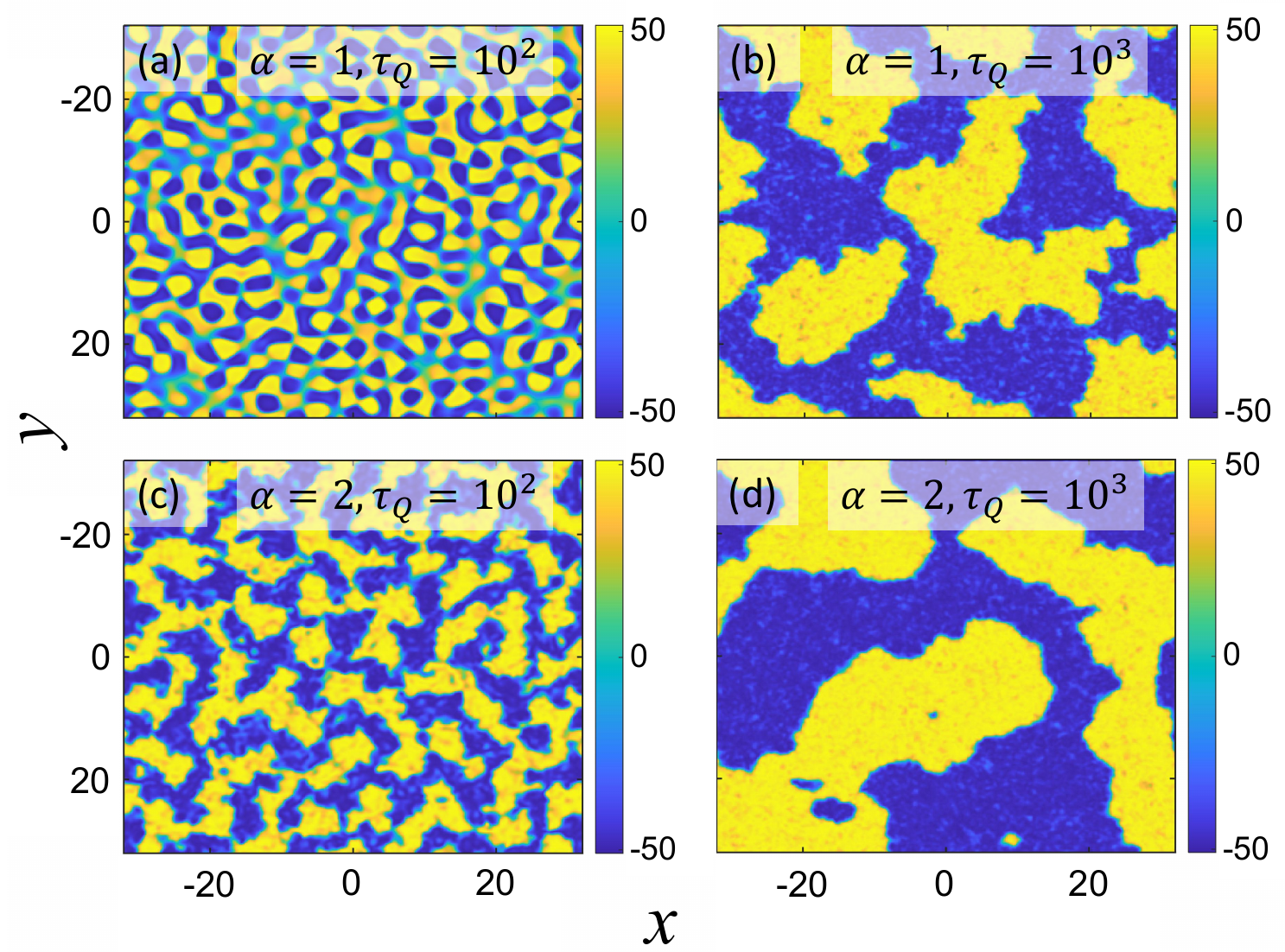}
  \caption{(Color online) The density difference $J(x,y)$ of two-component BECs, for varying quench time $\tau_{Q}$ and coefficient $\alpha$, exhibits a dynamical evolution that equally quenches to a common final parameter, i.e., $g_f=1.1g_c$.  }
\label{Fig_Dom_VS}
\end{figure}

In order to provide a qualitative description of the difference in wave-function between two-component BECs, we define the density difference as $J\left( x,y \right) = {\left| {{\psi _1}\left( x,y \right)} \right|^2} - {\left| {{\psi _2}\left( x,y \right)} \right|^2}$. The system is in the miscible phase when the density difference equals zero, i.e. $J\left( x,y \right)=0$; otherwise, it is in the immiscible phase.
The time evolution of the density difference $J(x,y)$ for a typical quench time $\tau_{Q}=10^{3}$ is shown in Fig.~\ref{Fig_Dom_Form}. Initially, the system was prepared in the miscible phase at $g_{12}^{i}=0.9g_{c}$ as depicted in Fig.~\ref{Fig_Dom_Form}(a). Subsequently, by linearly quenching the interaction $g_{12}(t)$ across the critical point according to Eq.\ref{Non_Quen_1}, it reaches a final value of $g_{12}^{f}=1.04g_{c}$ with a linear quench type, i.e ${\alpha}=1$. It is worth mentioning that before reaching $1.02g_c$, as shown in Fig.~\ref{Fig_Dom_Form}(a-c), the initial state almost stops evolving even when the parameter of the Hamiltonian enters the immiscible phase. However, due to the quenching process, there is a delay in experiencing the phase transition by the system. The formation of spontaneous domains occurs when deep within this immiscible phase, as seen in Fig.~\ref{Fig_Dom_Form}(d-f). These domains are found to be unstable during subsequent evolution due to excitation interactions within them. Furthermore, we observe that both domain density and delay time for phase transition strongly depend on quench time $\tau_Q$ and coefficient ${\alpha}$. In Fig.~\ref{Fig_Dom_VS}, we present density differences $J(x,y)$ at end of evolution for different values of ${\alpha}$ and $\tau_Q$. For smaller $\tau_Q$, changes occur more rapidly within Hamiltonian parameters leading to increased non-adiabaticity upon crossing critical point resulting in generation of more defects during subsequent evolution (see Fig.\ref{Fig_Dom_VS}(a,c)).

In our system, the presence of non-adiabatic defects corresponds to the existence of domains. As $\tau_{Q}$ increases, both the number of non-adiabatic defects and the time delay in phase transition decrease, leading to a more adiabatic evolution of the system (see Fig.~\ref{Fig_Dom_VS}(b,d)). It is anticipated that as $\tau_{Q} \to \infty$, both non-adiabatic defects and phase transition delays will vanish, resulting in a return to equilibrium dynamics. Furthermore, by comparing the density difference $J(x,y)$ between the upper row ($\alpha=1$) and lower row ($\alpha=2$) in Fig.~\ref{Fig_Dom_VS}, we have observed that a higher coefficient ${\alpha}$ effectively reduces the occurrence of defects, indicating that larger coefficients suppress non-adiabatic defects.

To quantitatively investigate the time delay of the phase transition, we define the order parameter as
\begin{equation}\label{Order_para}
\Delta J = \frac{1}{2} - \frac{1}{2N}\int\int{\left(\psi_{1}^{*}\psi_{2}+c.c.\right)}dxdy.
\end{equation}
 In the miscible phase, where the wave-function amplitudes for two-component BECs are nearly equal with a global phase difference, the overlap integration of  $\int\int{\psi_{1}^{*}\psi_{2}}dxdy$  approaches $N/2$. However, this overlap integration vanishes in the immiscible phase. Therefore, changes in $\Delta J$ can serve as an order parameter to characterize the MI phase transition.

\begin{figure*}[!htp]
  \centering\includegraphics[width=0.9\textwidth]{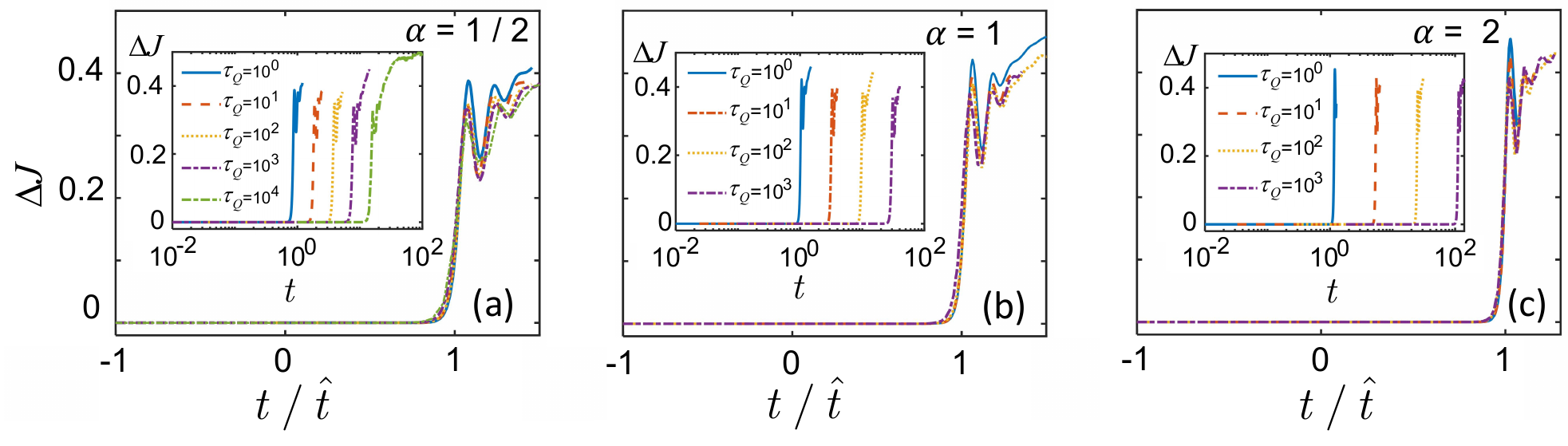}
  \caption{(Color online) Time evolution of the order parameter $\Delta J $ for different quench time $\tau_Q$  and quench coefficients ${\alpha}$, ${\alpha}$= 1/2 (a), ${\alpha}$ = 1 (b), and ${\alpha}$ = 2 (c). The critical point occurs at the $t/\hat{t}=0$ in our system.}
\label{Fig_Odp_Hom}
\end{figure*}

In Fig.~\ref{Fig_Odp_Hom}, we clearly demonstrate the temporal variation of $\Delta J$ for different quench times $\tau_{Q}$ and  coefficients ${\alpha}$. According to the KZM, the instantaneous state freezes at $-\hat{t}$ and remains unchanged until it passes over $+\hat{t}$. During the first adiabatic region and impulse region, the instantaneous state retains most of the information in the miscible phase, resulting in a zero value for the order parameter $\Delta J$. However, when crossing $+\hat{t}$, the instantaneous state starts evolving again but is no longer an eigenstate of Hamiltonian. Consequently, non-adiabatic defects occur as a result of a non-adiabatic process from freezing at $-\hat{t}$ to a symmetry breaking state after $+\hat{t}$, leading to a nonzero value for $\Delta J$.

In equilibrium processes, the order parameter becomes nonzero at phase transition from symmetry phase to symmetry breaking phase. In our non-equilibrium dynamics system, however, there is a delay in increasing order parameter $\Delta J$ when evolving at  $+\hat{t}$. This delay in phase transition strongly depends on quench time $\tau_{Q}$ (see the inset of the Fig.~\ref{Fig_Odp_Hom}). To quantitatively describe this delay in phase transition, we determined freezing time $\hat{t}$ when $\Delta J$ reaches a small nonzero value ($\delta J = 0.1$). Our simulations demonstrate that similar conclusions can be inferred for other ranges of $\delta J$, spanning from 0.05 to 0.35, due to the pronounced increase in $\Delta J$ near the freezing time $+\hat{t}$. The convergence of $\Delta J$ curves for different quench times suggests the existence of a scaling law between freezing time $\hat t$ and quench time $\tau_Q$, as depicted in Fig.~\ref{Fig_Odp_Hom}(a,b,c).

\begin{figure}[!htp]
  \centering\includegraphics[width=0.49\textwidth]{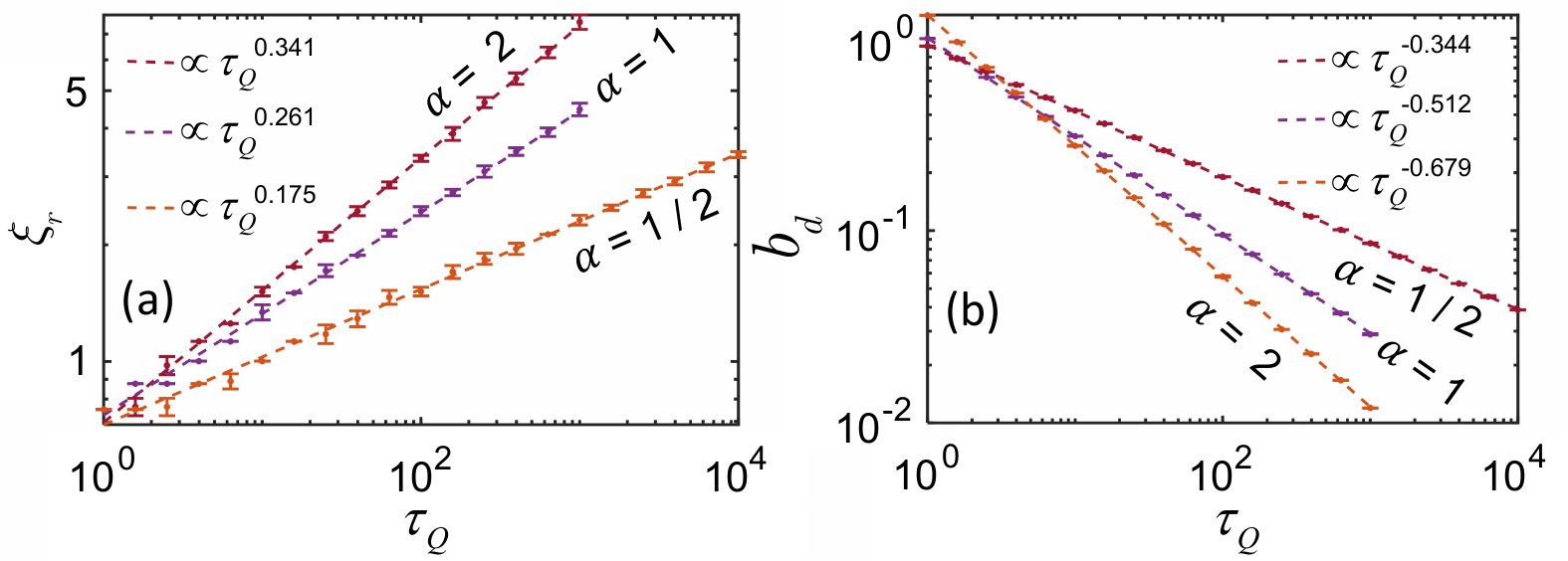}
  \caption{(Color online) . The temporal universal scaling of the phase transition delay and the spatially universal scaling of the domain for different quench times $\tau_{Q}$  and $\alpha$.
 The correlation length of domains is counted at the freeze time $+\hat{t}$ at which the instantaneous state restarts to evolve. Error bars correspond to the standard deviation of $10$ runs.  }
\label{Fig_Scaling_Hom}
\end{figure}

To investigate the relationship between phase transition delay and quench time, we define $b_d$ as
\begin{equation}\label{phase_delay}
{b_d}\sim \hat{\epsilon} \sim \left| {{g_{12}}\left( \hat{t} \right) - g_{12}^c} \right|\sim {\left( {\frac{\alpha}{{{\tau _Q}}}} \right)^{ - \frac{{\alpha}}{{1 + {\alpha}vz}}}}.
\end{equation}
 The universal scaling of phase transition delay $b_d$ with respect to quench time $\tau_{Q}$ is demonstrated in Fig.~\ref{Fig_Scaling_Hom}(b). In the case of linear quenching $(\alpha=1)$, the numerical scaling law for the phase transition delay $b_d$ in Fig.~\ref{Fig_Scaling_Hom}(b) is found to be approximately $0.512$, which closely matches the analytical scaling of $1/2$.
  The  value of $b_d$ decreases with increasing $\tau_Q$, suggesting a more adiabatic system behavior, ultimately vanishing when $\tau_Q\to \infty$.
  Moreover, a larger coefficient ${\alpha}$ can effectively reduce the delay time. The numerical scaling law of $b_d$ for $\alpha=2$ is approximately $0.679$, which closely aligns with the analytical result of $2/3$. This indicates a significant decreasing in the phase transition delay compared to linear quenching.
  The numerical scaling of $b_d$ for different ${\alpha}$ agrees well with the analytical results are summarized in Table~\ref{Critical_Exponent_Table}.

  \begin{figure}[!htp]
  \centering\includegraphics[width=0.48\textwidth]{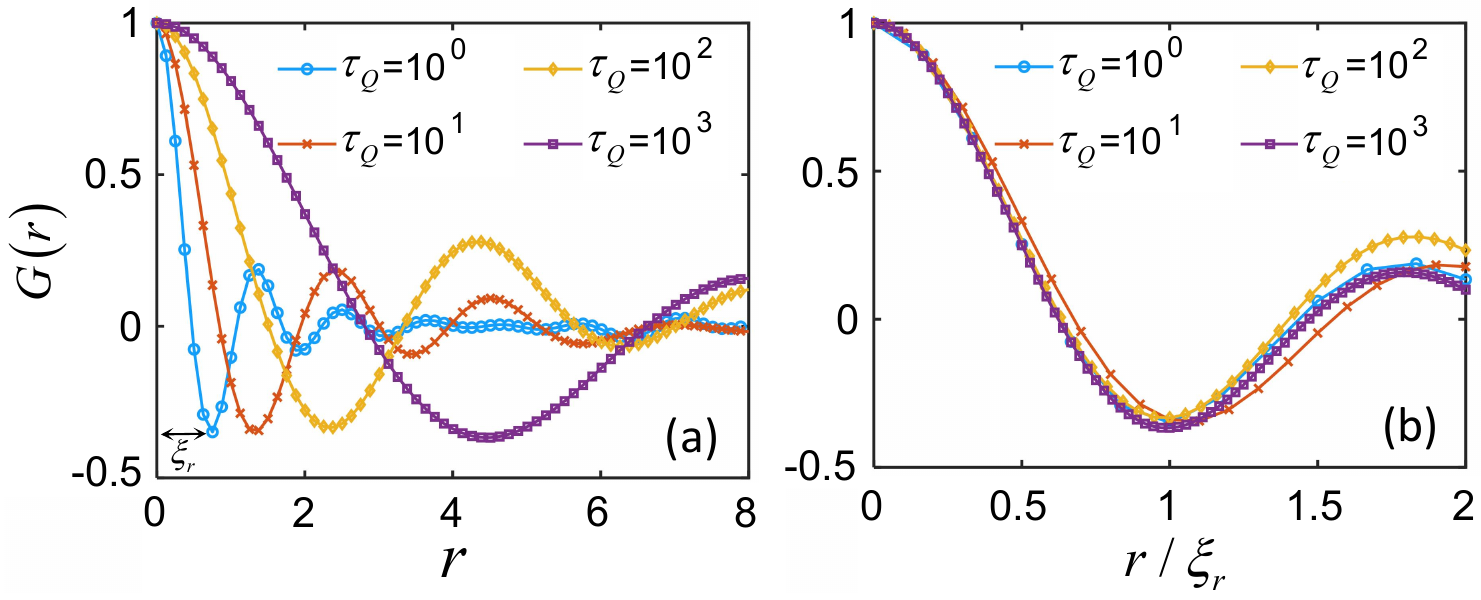}
  \caption{(Color online)  The spatial correlation functions $G(r)$ are illustrated in (a), showing examples of the functions for different quench times $\tau_Q$   as a function of distance $r$. In (b), it is demonstrated that the correlation functions $G(r)$ collapse onto a single curve when the distance is rescaled by the domain size $\xi_{r}$.}
\label{Fig_Corel_Hom}
\end{figure}

 To characterize domain correlation length, we analyze the density-density correlation function.
\begin{equation}\label{Den-Correl_func}
\begin{array}{l}
G\left( r \right) \equiv G\left( {x,y} \right)=\\
\int {\int {J (x^{'},y^{'} )J (x^{'}+x,y^{'}+y )dx^{'}dy^{'}}},{\rm{  }}r = \sqrt {{x^2} + {y^2}}
\end{array}
\end{equation}
In Fig.~\ref{Fig_Corel_Hom}, we present the correlation functions $G(r)$ for different quench times $\tau_{Q}$ in a linear quench dynamic. As the quench time increases, the domain size becomes larger, indicating a higher degree of particle correlation within the domain. The corresponding correlation length is denoted as $\xi_{r}$.
In this study, we determine $\xi_{r}$ by extracting it from the first minimum of the density-density correlation function, as depicted in Fig.~\ref{Fig_Corel_Hom}(a). Upon rescaling the distance by the domain size $\xi_{r}$, all plots of $G(r)$ for different $\tau_Q$ collapse onto a universal curve, as illustrated in  Fig.~\ref{Fig_Corel_Hom}(b),
indicating the existence of a universal correlation between the correlation length $\xi_{r}$ and quench time $\tau_Q$.
As the value of $\tau_Q$ increases, the domain size denoted by $\xi_{r}$ also increases, exhibiting a numerical scaling of 0.261 with respect to $\tau_Q$ for linear quenching, see Fig.\ref{Fig_Scaling_Hom}(a), this numerical scaling is in agreement with the analytical scaling of 1/4. Additionally, we observe that a larger quench coefficient $\alpha$ can further enhance the domain size, indicating that higher values of $\alpha$ can effectively suppress non-adiabatic defects. Specifically, for $\alpha$ = 2, the universal scaling between $\xi_{r}$ and $\tau_Q$ approaches 0.341, which is in close proximity to the analytical value of  $1/3$, and significantly larger than those obtained for $\alpha$ equal to 1 and 1/2. The numerical and analytical universal scalings of $\xi_{r}$ and $\tau_Q$ for different quench coefficients $\alpha$ are summarized in Table.~\ref{Critical_Exponent_Table}.

\begin{table}
\normalsize
\begin{center}
\begin{tabular}{|l|l|l|l|l|}
\hline
  Coefficient  & Theor $\xi_r$ & Nume $\xi_r$ & Theor $b_d$  & Nume $b_d$  \\ \hline

  $\alpha=1/2$  & 1/6 & 0.175 & 1/3  & 0.344 \\ \hline

  $\alpha=1$ & 1/4 & 0.261 & 1/2  & 0.512 \\ \hline

  $\alpha=2$ & 1/3 & 0.341 & 2/3 & 0.679 \\ \hline

\end{tabular}
\caption{The numerical and theoretical power laws of the $\xi_r$ and $b_d$ of the two-component BEC for different $\alpha$. }
\label{Critical_Exponent_Table}
\end{center}
\end{table}

\subsection{Inhomogeneous Case}
Below, we investigate the impact of a harmonic trap on non-equilibrium universal dynamics. In experiments, the harmonic trap is commonly used instead of a homogeneous trap, and its spatial inhomogeneity leads to non-uniformity in non-adiabatic defects. As such, this inhomogeneity can generally affect the universality of these defects. In this section, we examine the MI phase transition universal dynamics within a harmonic trap with for various quench coefficient $\alpha$. It is worth noting that space's inhomogeneity does not alter the systematic critical point; namely, $g_{12}^{c}$ remains independent of particle density.

\begin{figure*}[!htp]
  \centering\includegraphics[width=0.9\textwidth]{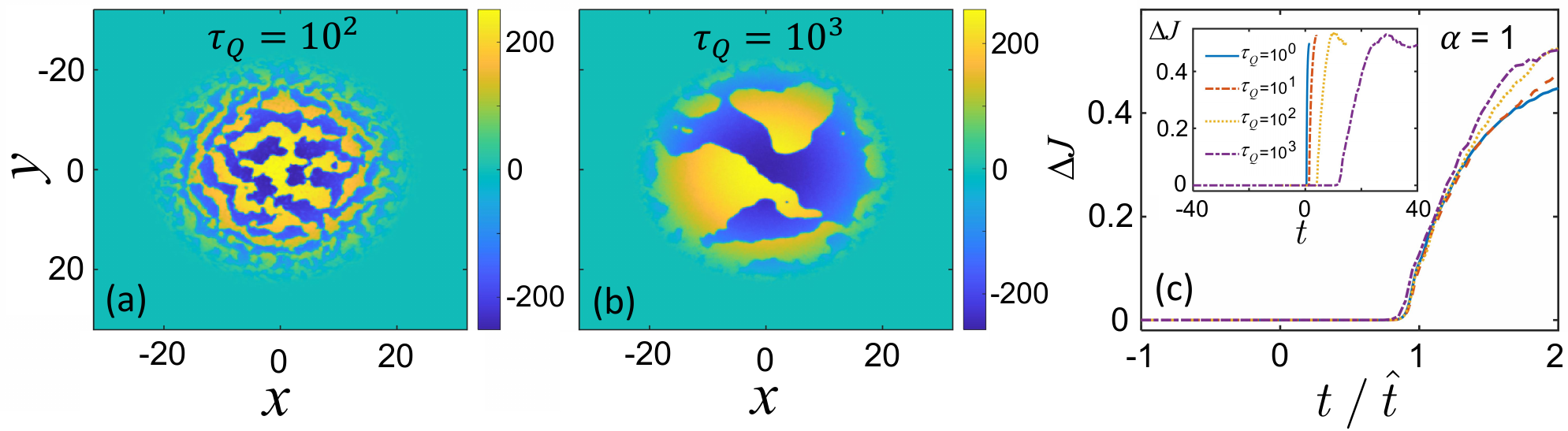}
  \caption{(Color online) (a,b)The final evolution of the density difference of two-component BECs for different  quenched dynamics with trap frequency $\omega^{2}=1$. (c) Time evolution of the order parameter $\Delta J$ for different quench time $\tau_Q$.  }
\label{Fig_InH_Dom_Odr}
\end{figure*}

To study non-equilibrium dynamics within a harmonic trap, we first prepare an initial state with Gaussian shape within the miscible phase and add noise to it before quenching interaction $g_{12}(t)$ across the critical point according to Eq.\ref{Non_Quen_1}.
The typical example of final density disparity $J(x,y)$ for the system confined within the trap is visualized in Fig.\ref{Fig_InH_Dom_Odr}(a). The domain structure prominently demonstrates a spatial density dependence resulting from the harmonic trap. Furthermore, upon generation of the domain, it exhibits instability due to mutual interactions, leading to an increment in domain correlation length towards equilibrium states. For larger $\tau_{Q}$ values, the instantaneous state at freeze time exhibits a higher correlation of spontaneous domains compared to smaller $\tau_{Q}$ values, as observed in Fig.\ref{Fig_InH_Dom_Odr}(b). This indicates that the instantaneous state will also approach the equilibrium state for larger quench time.

To quantitatively characterize the phase transition of MI, we also employ $\Delta J$ as the order parameter in a harmonic trap. In the miscible phase, when both wave-function of the two-component BECs exhibit a Gaussian shape for their instantaneous states, $\Delta J$ equals zero; otherwise, it indicates that the instantaneous state resides in the immiscible phase. In Fig.~\ref{Fig_InH_Dom_Odr}(c), we present the temporal variation of $\Delta J$ with linear quenching dynamics for different quench time, clearly illustrating the freezing and subsequent evolution of the instantaneous state in accordance with the KZM. In comparison to the homogeneous case, the presence of a harmonic trap effectively dampens the oscillation of $\Delta J$. The order parameter $\Delta J$ remains zero in the first adiabatic region (prior to $-\hat{t}$) and in the impulse region (between $-\hat{t}$ and $+\hat{t}$), where $t/\hat{t}=0$ represents the critical point. Following $+\hat{t}$, the system restarts its evolution and experiences a significant increase in $\Delta J$ within the deep immiscible phase. We define the freeze time $+\hat{t}$ as the moment when $\Delta J$ reaches a small nonzero value of $\delta J$ ($\delta J = 0.1$ in our simulation). The specific value of $\delta J$, ranging from 0.05 to 0.25, does not alter the universal scaling behavior of the freeze time or its associated phase transition delay.
 The plots of the order parameter $\Delta J$ for different quench times $\tau_Q$ exhibit a collapse onto a curve spanning from $-\hat{t}$ to $+\hat{t}$, indicating the presence of universal scaling between the freeze time and quench time, see the inset of Fig.~\ref{Fig_InH_Dom_Odr}(c).

\begin{figure}[!htp]
  \centering\includegraphics[width=0.49\textwidth]{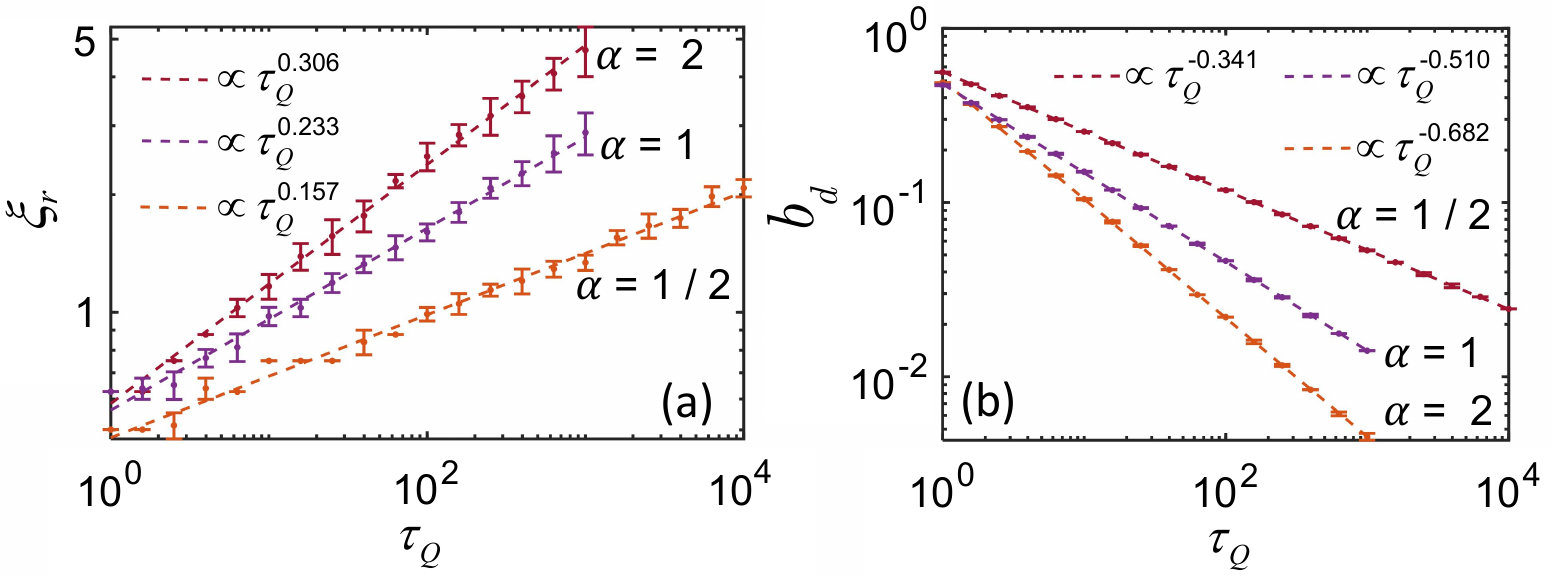}
  \caption{(Color online) . The temporal universal scaling of the phase transition delay and the spatially universal scaling of the domain for different quench times $\tau_{Q}$  for different coefficient $\alpha$.
 The correlation length of domains is counted at the freeze time at which the instantaneous state restarts to evolve. Error bars correspond to the standard deviation of 10 runs.  }
\label{Fig_Inh_Scaling}
\end{figure}

As mentioned above, we define the delay of the MI phase transition according to Eq.\ref{phase_delay}. In Fig.\ref{Fig_Inh_Scaling}(b), we present the temporal universal scaling of the phase transition delay in relation to the quench time for various coefficient $\alpha$.
As the quench time increases, the delay of the phase transition exhibits a power-law decay, indicating that the system will undergo a phase transition at the critical point as $\tau_Q \to \infty$. In this scenario, the non-equilibrium dynamics would converge to an equilibrium phase transition.
The scaling exponent between the phase transition delay $b_d$ and the quench time $\tau_Q$ in linear quenching is 0.510, which closes to the analytical results and  remains independent of the trap frequency.
It is noteworthy that the universal scaling of the phase transition delay $b_d$ in harmonic traps is akin to the homogeneous findings, due to the space-independent nature of the system's critical point.
The corresponding analytical and numerical universal scaling of $b_d$ for different trap frequencies and quench coefficients $\alpha$ are summarized in Table.\ref{Critical_Exponent_Table2}.

\begin{figure}[!htp]
  \centering\includegraphics[width=0.49\textwidth]{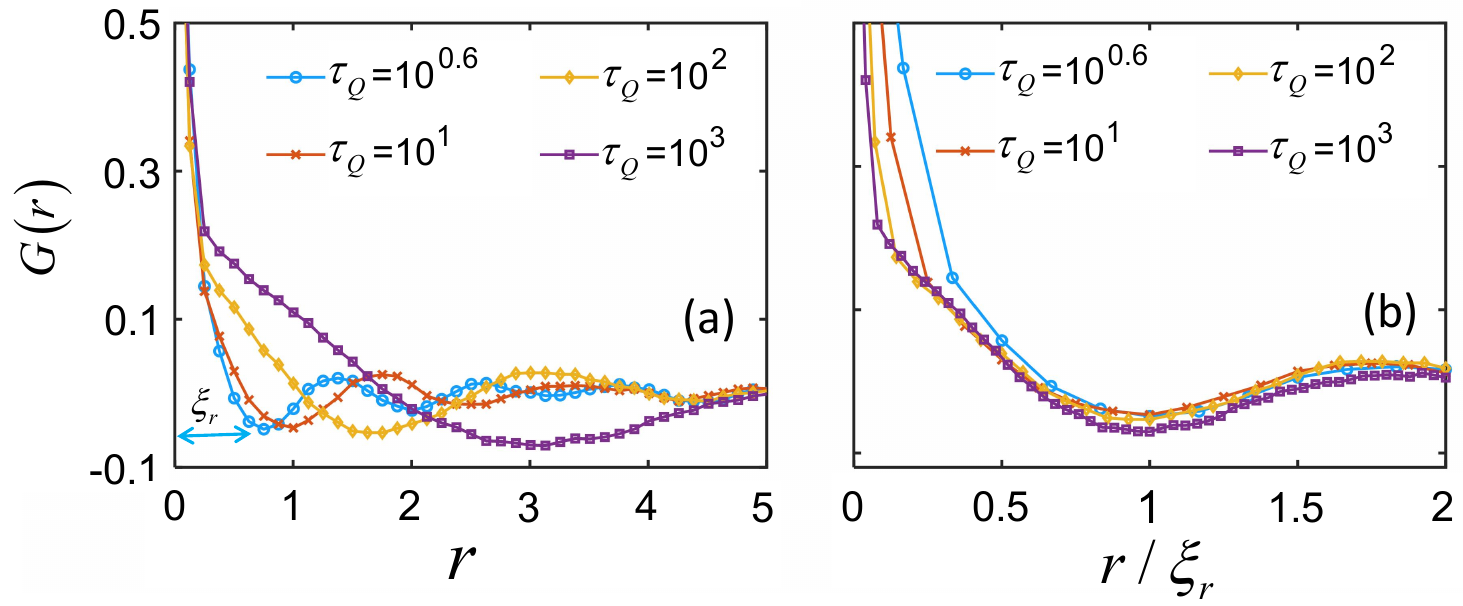}
  \caption{(Color online)  The spatial correlation functions $G(r)$ and their collapse with respect to the separated distance $r$ are observed in linear quenching. }
\label{Fig_Inh_Corel}
\end{figure}

In Fig.\ref{Fig_Inh_Corel}, we demonstrate the spatial correlation functions $G(r)$ for various quench times $\tau_Q$. The oscillation of $G(r)$ indicates the correlation length scale of the distribution of domains. We identify the position of the first minimum as the correlation length $\xi_{r}$. The $\xi_{r}$ increases with larger quench times, indicating a larger average area of domain defects.  The density distribution in $G(r)$ exhibits non-smoothness at the initial distances $r$ due to the density nonuniformity induced by the trap.
 To further investigate the universality, we represent the correlation functions $G(r)$ in the rescaled coordinate $r/\xi_{r}$ for different quench times. The correlation functions $G(r)$ exhibit a remarkable collapse to a linear trend, as illustrated in Fig.\ref{Fig_Inh_Corel}(b), indicating the presence of universality between $\xi_{r}$ and quench time $\tau_{Q}$.

In Fig.~\ref{Fig_Inh_Scaling}(a), we demonstrate the spatially universal scaling behavior of the domain for different quench times $\tau_{Q}$ and coefficient$\alpha$. The correlation length $\xi_r$ increases with the quench time $\tau_Q$, and when the quench time is sufficiently large, $\xi_r$ will reach a value of the same order of magnitude as the system length. The numerical universal scaling of the domain for linear quenching  is approximately 0.233, which is comparatively lower than the analytical scaling of $1/4$. This observation suggests that the presence of a harmonic trap can effectively reduce the correlation length.
The universal spatial scaling of domain for various quench times $\tau_{Q}$ and trap frequencies $\omega$ is summarized in the Table.~\ref{Critical_Exponent_Table2}.

\begin{table*}
\normalsize
\begin{center}
\begin{tabular}{|l|l|l|l|l|l|l|}
\hline
  Coefficient  & \makecell[tl]{Theor $\xi_r$, \\ $\omega^2=0$}    & \makecell[tl]{Nume $\xi_{r}$,\\ $\omega^2=0.5$}   & \makecell[tl] {Nume $\xi_{r}$,\\$\omega^2=1$}  &  \makecell[tl]{Theor $b_d$,  \\ $\omega^2=0$}  & \makecell[tl]{Nume $b_{d}$,\\$\omega^2=0.5$}  & \makecell[tl]{Nume $b_{d}$,\\$\omega^2=1$}  \\ \hline

  $\alpha=1/2$  & 1/6 & 0.170 & 0.157 & 1/3  & 0.340 & 0.341 \\ \hline

  $\alpha=1$ & 1/4 & 0.245 & 0.233 & 1/2  & 0.510 & 0.510 \\ \hline

  $\alpha=2$ & 1/3 & 0.313 & 0.306 & 2/3 & 0.678 & 0.682 \\ \hline

\end{tabular}
\caption{The numerical and theoretical power laws of the $\xi_r$ and $b_d$ of the two-component BEC for various trap frequency $\omega$ and quench coefficient $\alpha$. }
\label{Critical_Exponent_Table2}
\end{center}
\end{table*}

\section{Summary}

 In summary, we have investigated the universally critical dynamics dynamics of two-component Bose-Einstein condensates in a 2D system with two types of potential traps.
 By linearly and nonlinearly quenching the interaction strength across the critical point, we have studied domain formation and the universally critical dynamics of the miscible-immiscible phase transition.
 In a homogeneous trap, domain defects were homogeneously generated, while in a harmonic trap, domain formation depended on the density distribution.
 Unlike the domains formed in 1D systems, those formed in 2D systems were found to be unstable and tended to merge with each other during subsequent evolution.
 By calculating the correlation length at the freeze time, we observe that a larger nonlinear coefficient can suppress the number of defects, while the inhomogeneity of the trap can affect the universal scaling behavior of the correlation length.
 We find that the trap type has no influence on the scaling relationship between phase transition delay and quench time by varying trap frequency.
 Theoretical analysis and numerical simulations reveal a power-law scaling relation among domain correlation length, phase transition delay, and quench time.
 The obtained universal scaling from numerical simulations agrees well with our theoretical predictions.
 Our study confirms that the critical exponents of two-component BECs remain unaffected by dimensionality, and in 2D, defects are confined to specific domains.
 Furthermore, the nonlinear power-law quenching does not alter the critical exponent of the phase transition, thereby offering valuable insights into suppressing nonadiabatic excitations towards adiabaticity.

\begin{acknowledgments}
This work was supported by National Natural Science Foundation of China$($Grant No. 12305013 $)$,
the GuangDong Basic and Applied Basic Research Foundation $($Grant No. 2021A1515111015$)$,  the National Natural Science Foundation of China $($Grant Nos. 11904051, 12274077$)$, the Natural Science Foundation of Guang-dong Province $($Grant Nos. 2021A1515010214, 2023A1515010770$)$, the Special Funds for the Cultivation of Guangdong College Students' Scientific and Technological Innovation $($Grant No. pdjh2022a0538$)$, and the Research Fund of Guangdong-Hong Kong-Macao Joint Laboratory for Intelligent Micro-Nano Optoelectronic Technology $($Grant No. 2020B1212030010$)$.
\end{acknowledgments}

\bibliography{TD_Non_KZM_Ref}

\end{document}